\documentstyle[aps,prl,preprint]{revtex}
\draft
\begin{document}
\title{ Comment on the article by Bender et al, J. Phys. A. 35 (2002) L467
}
\author{C. R. Handy}
\address{Department of Physics \& Center for Theoretical Studies of 
Physical Systems, Clark Atlanta University, 
Atlanta, Georgia 30314}
\date{Received \today}
\maketitle
The recent Letter by Bender, Berry, and Mandilara (2002, BBM) 
presents some interesting symmetry arguments which enable one
to transform non-hermitian, PT invariant, (complex) polynomial 
potential hamiltonians, into secular equation representations 
with real coefficients. This achievement is claimed to ``demistify''
why these systems can admit real and/or complex eigenenergies.
Two approximations underly their arguments, and weaken the
implied significance of their work. The first is
that any practical numerical analysis of the secular equation
must involve a truncation of the system (i.e. reducing the full,
infinite dimensional hamiltonian matrix, to finite dimension).
What the research community wants is exact formulations that
can clarify under what conditions the eigenenergies of these
systems can be real.
The second ``approximation''
 is that they do not define any explicit conditions, 
with respect to the Hamiltonian parameters, for which real 
(approximate) eigenenergies to the secular equation, are to be expected.
 From these 
perspectives, their work is not any more significant than that of others
 they fail to cite. In particular, the work by Handy 
on the use of (real) nonnegative representations for the 
Schrodinger equation (2001a), as well as the generation of
converging bounds for both real and complex eigenenergies 
(2001b) are more relevant, and go further, than the intended focus
of the BBM analysis.
Specifically, within the Eigenvalue Moment Method (EMM) formulation,
one can generate {\it exact} (devoid of any truncation 
considerations) inequality constraints on the
discrete energy levels. 
It is through
this infinite hierarchy of exact relations that converging
bounds to the complex/real eigenenergies were demonstrated
for the $ix^3+iax$ potential (Handy 2001b), first studied by Delabaere and Trinh (2000),
as quoted by BBM. The extension to the
problem $x^4+ iax$ is a simple application of EMM. Comparable
extensions of EMM are given in the work of Yan and Handy (2001).
 The work of 
BBM may offer some future simplification on the real problem 
of understanding the existence of real eigenenergies for the
identified PT invariant, non-hermitian systems; however, as 
presented, it is still speculative, and unfairly misrepresents
its significance relative to other theoretical/computational
strategies published. 

Bender C M, Berry M V, and Mandilara A 2002   J. Phys. A: Math. Gen. 35 
L467

Delabaere E and Trinh D 2000 J. Phys. A: Math. Gen. 33 8771

Handy C R 2001a J. Phys. A: Math. Gen. 34 L271

Handy C R 2001b J. Phys. A: Math. Gen. 34 5065

Yan Z and Handy C R J. Phys. A: Math. Gen. 34 5593

\end{document}